\documentstyle[12pt,aps]{revtex}

\begin{document}
\draft
\title{On fermionic vacuum energy}
\author{She-Sheng Xue
}
\address{
ICRA, INFN  and
Physics Department, University of Rome ``La Sapienza", 00185 Rome, Italy
}

\date{December, 2000}


\maketitle

\centerline{xue@icra.it}

\begin{abstract}

We quantitatively compute the rate and spectrum of spontaneous photon emissions, caused by the vacuum decaying when an external magnetic field is introduced. The spectrum of spontaneous photon emissions shows the energy-momentum-dependency in the low-energy region and in the high energy region, is respectively similar to the Rayleigh-Jeans part and the Wien part of the spectrum of the black-body radiation. However, the energy-momentum of spontaneous photon emissions is quantized, analogous to the Wigner spectrum. This provides quantitative results for experimental testes.

\end{abstract}


\narrowtext

The vacuum has a very rich physical content in the context of relativistic quantum field theories. It consists of extremely large number of virtual particles and anti-particles. The quantum fluctuations of the vacuum are creations and annihilations of these virtual particles and anti-particles in all possible energy-range. As a consequence of the quantum fluctuations, the vacuum energy does not vanish. The vacuum energy (the zero-point energy) due to the quantum fluctuations of electromagnetic fields in the vacuum is positive, while the vacuum energy due to the quantum fluctuations of fermion fields in the vacuum is negative. The Casimir effect\cite{casimir} shows that the positive vacuum energy attributed to the quantum fluctuations of electromagnetic fields in the vacuum has a macroscopic effect. The effect proposed in ref.\cite{xue01} shows that the negative vacuum energy attributed to the quantum fluctuations of fermionic fields in the vacuum can possibly have a macroscopic effect as well. An external magnetic field induces the vacuum decaying, which releases the vacuum energy of, 
\begin{equation}
\Delta{\cal E}=8{\alpha \over3\pi}B^2V, \hskip0.5cm \alpha={1\over137},
\label{delta2}
\end{equation}
where $B$ is the strength of the magnetic field and $V$ is the volume of the space where the magnetic field is created.
If the vacuum state decays and the vacuum energy (\ref{delta2}) is released, the phenomena and effects that could occur are following: (i) The vacuum acts as a paramagnetic medium that effectively screens the strength of the external magnetic field $B$ to a smaller value $B'<B$ for the total energy-density being,
\begin{equation}
{1\over2}B'^2={1\over2}B^2-8{\alpha \over3\pi}B^2;\hskip0.5cm B'=B\sqrt{1-{16\over3\pi}\alpha}.
\label{hed}
\end{equation}
This phenomenon of paramagnetic screening could be possibly checked by appropriate experiments of precisely measuring the magnetic field strength. (ii) The vacuum-energy fluctuations could lead to the emission of neutrino and anti-neutrino pairs from the vacuum, since they are almost massless, which however is almost impossible for an experimental test.
(iii) Photons are spontaneously emitted analogously to the spontaneous photon emissions for electrons at high-energy levels decaying to low-energy levels in the atomic physics. This phenomenon should be possibly detected if any photons are emitted when the magnetic field $B$ is turned on. 
In this note, we present the quantitative computations of the rate and spectrum of spontaneous photon emissions. This provides quantitative result for experimental testes.   

For simplifying computations of such a transition amplitude at tree-level, we chose a particular Lorentz frame where the initial fermion state is at rest, the magnetic field $B$ is in $\hat z$-direction and the electric charge $e$ is renormalized. The initial state $\psi_i^{(-)}$ for $B(t)=0, t<0$ and final state $\psi_f^{(-)}$ for $B(t)=B, t>\Delta\tau$ of negative energy solutions are given by    
\begin{eqnarray}
\psi_i^{(-)}&=&\left({1\over V}{m\over E_i}\right)^{1\over2}e^{iE_it}\left(\matrix{0\cr\chi^\alpha}\right);\hskip0.3cm \psi_f^{(-)}=\left({1\over V}{m\over E_f}\right)^{1\over2}c_ne^{-{\xi^2\over2}}H_n(\xi)e^{iE_ft-p_y^fy-p_z^fz}\left(\matrix{0\cr\chi^\alpha}\right)\nonumber\\
E_i&=&m,\hskip0.5cm  E_f=\sqrt{m^2+(p^f_z)^2+eB(2n+1-\alpha)},
\label{ifstates}
\end{eqnarray}
where the spinor $\chi^\alpha$: $\sigma_z\chi^\alpha=\alpha\chi^\alpha$ for the helicity $\alpha=\pm 1$, $\xi=\sqrt{eB}(x-{p^f_y\over eB})$, $H_n(\xi)$ is the Hermite polynomial and $c_n=1/(2^{n\over2}\sqrt{n!}\pi^{1\over2})$\footnote{These negative energy
solutions can be obtained by the charge conjugation of corresponding positive energy-solutions.}. The probability of spontaneous photon emissions is related to the amplitude 
$|\epsilon_\mu^\beta J^\mu(k)|^2$\cite{iz}, where $\epsilon_\mu^\beta$ is the transverse polarization vector of photons emitted and 
\begin{equation}
J^\mu(k)=e\int d^4xe^{-ikx}\bar\psi_f^{(-)}\gamma^\mu\psi_i^{(-)},
\label{j}
\end{equation}
where ``$k$'' is photon's energy-momentum. Integrating variables $t,y$ and $z$ gives $\delta$-functions for energy-momentum conservations. Armed with eq.(7.376) in \cite{gr}, we integrate variable $x$,
\begin{equation}
\int dxe^{-ik_xx}e^{-{\xi^2\over2}}H_n(\xi)=(-i)^n({2\pi\over eB})^{1\over2}e^{-ik_x{p^f_y\over eB}}H_n({k_x\over \sqrt{eB}})e^{-{k_x^2\over 2eB}}.
\label{j}
\end{equation}
The computation of the amplitude $|\epsilon_\mu^\beta J^\mu(k)|^2$ is straightforward in the spinor space. Taking average over helicities of initial states, summing over all final states $p^f_z,p^f_y$ and $n$ with degeneracy $|e|BS/(2\pi)$,
as well as the polarizations of photons emitted, we obtain,
\begin{eqnarray}
{1\over2}\sum_\beta|\epsilon_\mu^\beta J^\mu(k)|^2 &=& e^2e^{-{k_x^2\over eB}}\sum_{n=1}^\infty{1\over2^nn!\pi}H^2_n({k_x\over \sqrt{eB}})(2\pi)\Delta\tau\delta[\omega_k-E^n+m]\left({m\over E^n_f}\right)\nonumber\\
E^n_f&=&\sqrt{m^2+(k_z)^2+2eBn}
\label{re}
\end{eqnarray}
where $\omega_k=|k|$ is the photon energy and $\delta$-function for the energy-conservation. The term corresponding to $n=0$ has been dropped for energy-momentum conservations, since the $\delta$-function $\delta[\omega_k-E^{n=0}_f+m]$ only gives solution $|k|\equiv 0$. Because the problem is axial symmetric w.r.t. $\hat z$-direction, we can make substitutions $k^2_x\rightarrow k_\perp^2=k_x^2+k_y^2$, $k_x\rightarrow |k_\perp|$ and define $k_z=k_\parallel$ in eq.(\ref{re}). The $\delta$-function in eq.(\ref{re}) can be given as 
\begin{equation}
\delta[\omega_k-E^n_f+m]=\left({m+|k|\over eB}\right)\delta_{n,n_\circ},\hskip0.2cm 
n_\circ={k_\perp^2+2|k|m\over 2eB},\hskip0.2cm E^{n_\circ}_f=m +|k|,
\label{n}
\end{equation}
where $n_\circ=1,2,3,\cdot\cdot\cdot$, indicating the energy-momentum of emitted photons is quantized. As a result, eq.(\ref{re}) is,
\begin{equation}
{1\over2\Delta\tau}\sum_\beta|\epsilon_\mu^\beta J^\mu(k)|^2 = {e^2\over2^{n_\circ} n_\circ !\pi}e^{-{k_\perp^2\over eB}}H^2_{n_\circ}({|k_\perp |\over \sqrt{eB}})(2\pi)\sum_f(Q^2_f)\left({m_f\over eB}\right),
\label{fre}
\end{equation}
where $\sum_f$ is over all flavors of charged fermions. We find that eq.(\ref{fre}) does not explicitly depend on $k_\parallel$. For given $n_\circ=n_\circ(k_\perp,k_\parallel)$ in (\ref{n}), eq.(\ref{fre}) (the probability of photon emissions) is very small for $|k_\parallel|\gg |k_\perp|$, because the polynomial $H^2_{n_\circ}({|k_\perp |\over \sqrt{eB}})$ in eq.(\ref{fre}) is very small for small values of $|k_\perp|$. Thus, the most probability of photon 
emissions occurs for $|k_\perp|\gg |k_\parallel|$, indicating most emitted photons are near in the plane perpendicular to the magnetic field $B$. This is analogous to the phenomenon of synchrotron radiation. 

The probability $p_{n_\gamma}$ corresponding to the emission of $n_\gamma$ photons, when neither the momentum nor the polarizations are observed, is given by the Poisson distribution\cite{iz},
\begin{equation}
p_{n_\gamma} = {\bar n_{\gamma}\over n_{\gamma}!}e^{-\bar n_{\gamma}},
\label{pn}
\end{equation}
where $\bar n_{\gamma}$ is defined by
\begin{equation}
\bar n_{\gamma} =\int d\tilde k{1\over2}\sum_\beta|\epsilon_\mu^\beta J^\mu(k)|^2,\hskip0.3cm \int d\tilde k\equiv \int{d^3k\over (2\pi)^32\omega_k}\simeq\int\left({dk_\parallel dk_\perp\over (2\pi)^22}\right)_{\omega_k\simeq |k_\perp|},
\label{np}
\end{equation}
which is actually the average number of emitted photons, $\bar n_{\gamma}=\sum_\circ^\infty n_{\gamma}p_{n_\gamma}$. The number- and energy-spectrum of spontaneous photon emissions in a phase space element $d\tilde k$ and a unit of time are given by
\begin{equation}
{d\bar n_{\gamma}\over d\tilde k} ={1\over2\Delta\tau}\sum_\beta|\epsilon_\mu^\beta J^\mu(k)|^2,\hskip0.5cm 
{d\bar \epsilon_{\gamma}\over d\tilde k} ={1\over2\Delta\tau}\omega_k\sum_\beta|\epsilon_\mu^\beta J^\mu(k)|^2.
\label{snp}
\end{equation} 

We estimate that $\sqrt{eB}\simeq 0.244$eV for $B=10^5G$ achieved in the laboratory today\cite{magnet} and $\sqrt{eB}\simeq 24$KeV for $B\simeq 10^{15}G$ around neutron stars, i.e., $\sqrt{eB}\ll m_f$. We consider the limit for emitted photons whose energy-momentum $|k|\ll m_f$. From eq.(\ref{n}), we have $n_\circ=|k|m_f/(eB)$, i.e.,
$|k|=\omega_k=n_\circ eB/m_f$, showing the energy of emitted photons is quantized in the unit $eB/m_f$. As seen from eq.(\ref{fre}), the probability is exponentially suppressed for large values of $k_\perp^2/ eB$ and also suppressed by $1/(2^{n_\circ} n_\circ !)$ for large values of $n_\circ$. As a consequence, most photons emitted should have the momentum $
|k_\perp|\sim eB$ in the infrared region and quantized $|k_\perp|\simeq\omega_k=n_\circ eB/m_f$ for the small values of $n_\circ$.  
For $n_\circ=1, |k_\perp|\simeq |k|=\omega_k=eB/m_f$ and $H_1(x)=2x$, we have the rate, 
\begin{equation}
{1\over2\Delta\tau}\sum_\beta|\epsilon_\mu^\beta J^\mu(k)|^2=4e^2e^{-{k_\perp^2\over eB}}k_\perp^2\sum_f(Q_f^2){m_f\over (eB)^2}.
\label{pre1}
\end{equation}
for the number- and energy- spectrum of spontaneous photon emissions. This spectrum shows $k_\perp^2$-dependency in the low-energy region and $\exp(-{k_\perp^2\over eB})$-dependence in the high energy region, respectively similar to the Rayleigh-Jeans part and the Wien part of the spectrum of the black-body   
radiation. However, the energy-momentum $|k_\perp|\simeq \omega_k$ is quantized, in this sense, it is more analogous to the Wigner spectrum of the distribution of discrete energy-levels of atoms and nuclei. 

Although it seems very difficult to experimentally test these effects, as discussed in ref.\cite{xue01}, we still expect a sophisticate experiment in near future to verify the phenomena and effects of the vacuum-energy releasing via paramagnetic screening and spontaneous photon emissions induced by the external magnetic field. This is important for the understanding of the fermion structure of the vacuum of relativistic quantum field theories and any possibly prospective applications\cite{application}.



\begin{references}
\bibitem{application}
The right to use this idea (as an intelligent product) for any applications is preserved by the author.
\end{references}
\end{document}